\documentclass[]{spie}  %>>> use for US letter paper
\usepackage{amsmath,amsfonts,amssymb}
\usepackage{graphicx}
\usepackage[colorlinks=true]{hyperref}

\def\arcmin{$^{\prime}$}
\def\arcsec{$^{\prime\prime}$}

\title{Development of the Arizona Robotic Telescope Network}

\author[a]{Benjamin J. Weiner}
\author[b]{David Sand}
\author[c]{Paul Gabor}
\author[b]{Chris Johnson}
\author[b]{Scott Swindell}
\author[d]{Petr Kub\'anek}
\author[b]{Victor Gasho}
\author[c]{Taras Golota}
\author[b]{Buell Jannuzi}
\author[b]{Peter Milne}
\author[b]{Nathan Smith}
\author[b]{Dennis Zaritsky}
\affil[a]{MMT/Steward Observatory, 933 N. Cherry St., Tucson, AZ 85721}
\affil[b]{Steward Observatory, 933 N. Cherry St., Tucson, AZ 85721}
\affil[c]{Vatican Observatory, 933 N. Cherry St., Tucson, AZ 85721}
\affil[d]{Large Binocular Telescope Observatory, 933 N. Cherry St., Tucson, AZ 85721}

\begin{document} 
\maketitle

\begin{abstract}
The Arizona Robotic Telescope Network (ARTN) project is a long term effort to
develop a system of telescopes to carry out a flexible program of PI
observing, survey projects, and time domain astrophysics including
monitoring, rapid response, and transient/target-of-opportunity followup. Steward
Observatory operates and shares in several 1-3m class telescopes with
quality sites and instrumentation, largely operated in classical
modes. Science programs suited to these telescopes are limited by
scheduling flexibility and available observer person-power. Our
goal is to adapt these facilities for multiple co-existing queued
programs, interrupt capability, remote/robotic operation, and delivery
of reduced data. In the long term, planning for the LSST era, we
envision an automated system coordinating across multiple telescopes
and sites, where alerts can trigger followup, classification, and
triggering of further observations if required, such as followup
imaging that can trigger spectroscopy. We are updating telescope
control systems and software to implement this system in stages,
beginning with the Kuiper 61'' and Vatican Observatory 1.8-m
telescopes. The Kuiper 61'' and its Mont4K camera can now be controlled and
queue-scheduled by the RTS2 observatory control software, and operated
from a remote room at Steward.  We discuss science and technical
requirements for ARTN, and some of the challenges in adapting
heterogenous legacy facilities, scheduling, data pipelines, and
maintaining capabilities for a diverse user base.
\end{abstract}


\section{Introduction}
\label{sec:intro}  % \label{} allows reference to this section

Astronomy is entering a new era of time domain science and large survey datasets, with the Large Synoptic Survey Telescope (LSST) national facility coming online in a few years, preceded by time-domain surveys such as the Palomar Transient Factory and Zwicky Transient Factory, large imaging programs for dark energy surveys, and time-critical events such as gamma-ray bursts and gravitational waves.  These, especially LSST, will produce a flood of candidate events including supernovae, stellar outbursts, flaring variability from accretion onto supermassive black holes, and yet-uncharacterized transient events.

In the era of LSST our discovery potential will be limited by our ability to follow up and characterize interesting objects found in the time-domain as variable sources, or in the deep imaging survey.  LSST will generate a vast publicly released stream of transient alerts, and even with filtering based on astrophysical knowledge and machine learning (e.g. the ANTARES event broker\cite{Saha16, Narayan18}), there will be many more events than the community is ready to observe. Follow-up observations including both initial characterization and monitoring of transients will be especially important in the early years of LSST operation to provide feedback into classification algorithms for event brokers. However, the classical mode in which most telescopes are now scheduled, often months ahead, severely limits the ability to trigger time sensitive observations, and makes it difficult to carry out monitoring projects due to limits on time and number of people to observe.  For example, in current telescope scheduling it is difficult to trigger an observation a day after an event, but it is also difficult to take an observation for one hour per night for a week, or one night every week for two months.

Our goal with the Arizona Robotic Telescope Network is to develop a network of telescopes that is flexibly scheduled and has multiple capabilities, including photometry, spectroscopy, wide-field optical imaging, rapid response, and monitoring and survey capabilities. We are updating and integrating multiple small to medium size telescopes ($\sim1-3$ meters) on sites operated by Steward Observatory, the Vatican Observatory, and our partners. While there are other telescope networks in planning and operation, our approach is distinct in that we are: 1. upgrading existing facilities at well developed, good sites; 2. planning for heterogeneous systems and capabilities to handle multiple science goals; 3. envisioning the network as a system that can ingest a variety of scientific programs and quickly respond to requests for observations; and 4. developing some level of autonomy, where the network could, for example, take and process follow-up imaging and depending on the result, trigger a spectroscopic observation.  This is an ambitious project and our plan is to work toward it in stages, developing remote and autonomous observing and data processing one site and instrument at a time.

In the first stages of this project we are upgrading Steward Observatory's Kuiper 61'' telescope and preparing it for remote, queue-scheduled, and autonomous operation with the RTS2 control system\cite{Kubanek08, Kubanek16} (Remote Telescope System, {\tt https://rts2.org/}); and upgrading the Vatican Advanced Technology Telescope (VATT) for similar modes of operation \cite{Swindell17}.  Future work on ARTN will include: a more friendly interface for users/PIs;  a telescope network manager to distribute operations among heterogeneous telescopes; upgrading Steward's Bok 90'' telescope for remote, queue, and autonomous operation; and defining an interface to allow observation requests to outside observatories. Here we describe some of the science cases for ARTN and discuss progress on the Kuiper 61'', which can now be queue-scheduled and controlled by RTS2.

\section{Scientific Motivation}

Some of the science cases that motivate ARTN are listed here, with a focus on transient and time-domain astronomy. This is by no means an exhaustive list.

Transient followup and characterization: Astronomical transients include violent events such as supernovae, gamma-ray bursts, tidal disruption of stars accreted into supermassive black holes, and gravitational-wave events from black hole and neutron star mergers.  Each of these is critical to observe at early times in its evolution. In addition, LSST will detect many other transients such as massive star outbursts, supernova precursors, and poorly studied phenomena such as kilonovae, and time-sampling and monitoring will be needed to classify these events. Telescopes smaller than LSST that have more flexible scheduling will be vitally important to follow their time evolution.

$\bullet$ Supernovae (exploding massive stars, or white dwarfs in binaries) are key to the measurement of dark energy, yet there are many unknowns about the progenitors, explosion mechanism, and extinction by circumstellar or interstellar dust. Supernovae evolve on the timescale of days, and require both quick followup and monitoring over days or months, to measure the light curve as it rises to peak and slowly declines, for understanding and calibrating the SN brightness relation; classifying types of supernova and understanding the different progenitor populations; and probing the physics of supernovae and their surrounding medium.  
%[Nathan Smith, Peter Milne]

$\bullet$ Gamma-ray bursts are believed to arise from the collapse of very massive stars, or the mergers of compact objects (white dwarfs, neutron stars, black holes). They probe star formation in the early universe, the fates of massive stars, and the physics of compact objects. These extreme events are highly beamed, detected in gamma-rays and X-rays by satellites, and localized by followup of the optical counterpart, when it exists, within minutes to hours\cite{Fong16}. Triggering followup of GRBs in less than a night (sometimes less than an hour) is necessary to measure the decay of the light curve, which probes the GRB physics and is needed to measure the redshift for the most distant GRBs.  
%[Wenfai Fong, Nathan Smith?]

$\bullet$ Gravitational wave events from black hole and neutron star mergers are an exciting field that is now becoming reality with detections from LIGO, the first detection of a kilonova from a neutron star-neutron star merger\cite{Soares-Santos17, Valenti17, Kilpatrick17}, and plans to build a space GW detector for supermassive BH mergers. NS-NS mergers have now been shown to have a detectable optical counterpart, while BH-BH mergers may not have an electromagnetic signature. The duration of an optical transient is likely to be extremely short (minutes-hours) and current GW detectors do not localize the event to a small area, so it will typically be necessary to pre-image galaxies likely to host an event, and then rapidly image possible hosts when an event goes off and process them to detect potential counterparts.
% [Wenfai Fong]

Monitoring campaigns: Another aspect of time-domain astronomy is carrying out programs that are not necessarily rapid-response, but require observations over a period of time of one or a class of objects. These can be difficult to carry out in a classical observing mode but are more easily accommodated on a queue-scheduled telescope, and are easier for the users when the telescope can be operated remotely or autonomously.

$\bullet$ Masses of supermassive black holes from reverberation mapping: Galaxies host SMBHs at their centers, which can power quasars and galactic nuclei when active, and have a tight relation between BH mass and galaxy mass, suggesting that the black hole and galaxy regulate evolution in some way. To constrain the evolution we estimate black hole masses in quasars and AGN at high redshift from emission line luminosities and velocity widths. These relations are calibrated through measuring BH masses with reverberation mapping, measuring the time delay between variations in the continuum luminosity and the emission line luminosity to estimate a radius and velocity. This reverberation mapping technique requires monitoring in photometry and in spectra of the emission lines over typical $\sim 1$ month time delays\cite{Grier17}. With a large photometric time survey like LSST, it will be possible to do opportunistic RM by watching for quasars or AGN that change strongly in brightness, and triggering spectroscopic monitoring over weeks to months to catch that change echoed in the emission lines. 
%[Ian McGreer, Xiaohui Fan]

$\bullet$ Late time imaging and spectroscopy of supernovae: By measuring the brightness decay and evolution in spectral features of supernovae  continuing long after the explosion (weeks to months or even years), we can probe the dust and gas around the supernova, as its ejecta expand outward and encounter circumstellar material\cite{Mauerhan14,Andrews18}. For massive-star supernovae, much of the circumstellar material was ejected from the progenitor star itself, allowing us to probe the physics of the star before the explosion. For white-dwarf SN Ia, critical to measuring dark energy, but whose progenitor channels are poorly constrained, the late-time evolution of the light curve gives clues to the ejecta and the physics of supernovae\cite{Milne15}.
%[Peter Milne, Nathan Smith]

$\bullet$ Solar system objects: These include studies of outer solar system minor planets, asteroid and comet families, and Near Earth Objects or Potentially Hazardous Asteroids.  The latter are of high interest for both astronomical and planetary defense reasons, and astronomers at the University of Arizona Lunar \& Planetary Laboratory, operating the Catalina Sky Survey, use multiple telescopes to discover, measure orbits for, and re-acquire NEOs.  Since NEO distance and magnitudes vary quickly with time, flexible scheduling on telescopes larger than the discovery telescope is useful. This science case places an additional technical requirement on both the telescope and the scheduler: they should be able to track, point, and schedule objects with non-sidereal motions. Scheduling is a common problem if a system is built on a data model that assumes a fixed or negligibly changing RA/Dec.

$\bullet$ Space situational awareness: Space debris and the need to monitor for satellite mishaps and possible collisions are an increasing area of concern for all satellite operators. Observations of space objects for photometric characterization are also an active research area. Flexible scheduling is necessary since the objects move rapidly and space events occur on short timescales. Observing Earth-orbiting satellites raises similar issues to solar system objects, but places requirements on non-sidereal tracking rates, response times, etc. that can be an order of magnitude faster, even in high earth orbit.

In addition, flexibly scheduled telescopes enable many survey science programs, including large surveys and non-time-domain projects, that are difficult to do currently due to scheduling or people-power limits, such as constraints on how often individuals can be physically present to operate the telescope.

\section{System Architecture}

The Arizona Robotic Telescope Network is envisioned to bring flexible
scheduling and remote or autonomous operations to a number of the small and medium telescopes (2.3-m and smaller) operated by Steward Observatory for the University of Arizona Observatories, and operated by its partners (e.g. the Vatican Observatory). These are a heterogeneous set of telescopes with a range of instrumentation capabilities and serve a diverse range of user interests.  In addition, it is necessary to preserve classical observing capabilities at some or most of these telescopes to accommodate a variety of uses, including specialized instrumentation or observation requirements, educational uses, instrumentation development, observer and student training, and so on.  

For these reasons, the ARTN architecture has to be distributed and relatively flexible, since it cannot assume that installations are similar or that an individual observatory will always be available. It needs to support a diverse range of capabilities. Scientifically and operationally, this is also a potential strength; for example individual telescopes could be set up for wide field imaging, smaller field dedicated imaging, and spectroscopy, and observations taken at one telescope could, given automated processing, trigger followup at another telescope.

\begin{figure} [ht]
   \begin{center}
   \includegraphics[width=15cm]{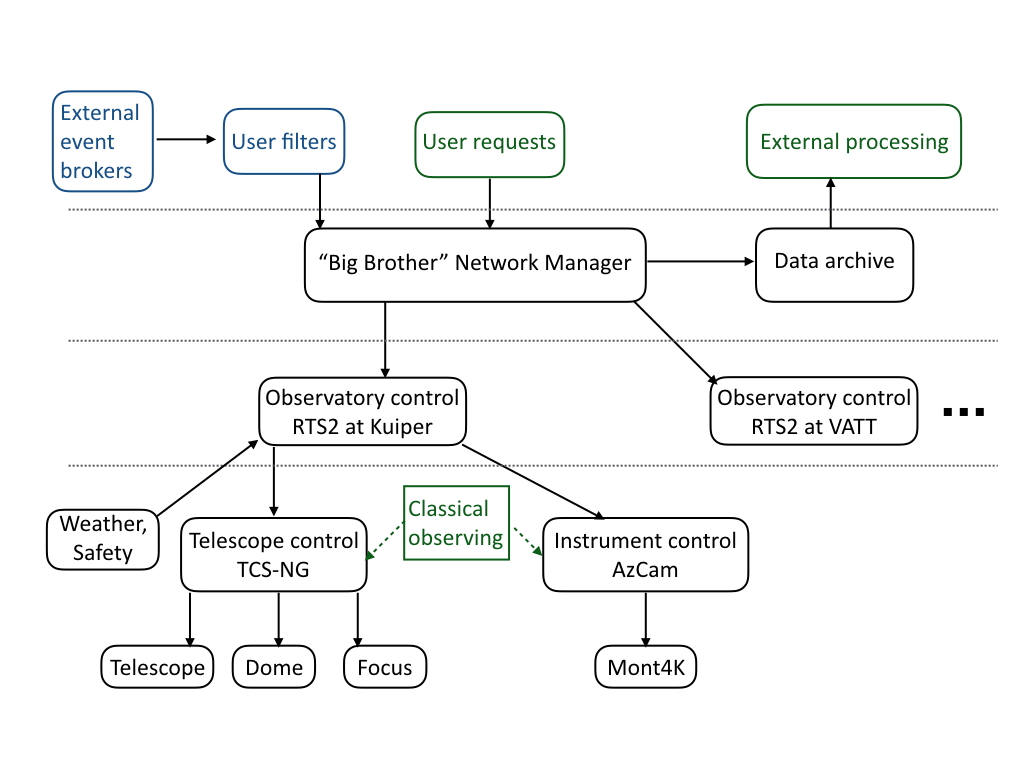}
  \end{center}
   \caption[] 
%>>>> use \label inside caption to get Fig. number with \ref{}
   { \label{fig:layers} 
A block diagram of the layers of control for ARTN network architecture. At the top layer, input comes from users and filtered alert streams from event brokers or other sources of triggers, and data is returned to users. At the network control layer, ``Big Brother'' manages the network and communicates to individual observatories. At the observatory layer, instances of RTS2 schedule and operate each telescope, communicating to the TCS and the instrument computers, which command the telescope, dome, and instrument.  Data and status reporting flow back up to the top level.
}
  \end{figure}

A sketch of a layered architecture for ARTN is shown in Figure \ref{fig:layers}. The uppermost layer is user responsibilities external to ARTN, in which decisions must be taken by the science users of the facility.  Here, users make observation requests individually, or filter alert streams from event brokers and turn these into observation requests programatically. Users also receive data taken by the system and apply their processing algorithms to it -- which could then generate further observation requests.

The top layer of ARTN is the network manager, aka ``Big Brother.'' 
The network manager communicates to individual observatory control systems, here shown as instances of RTS2, which communicates through an API of ASCII commands. In practice, it is likely that Steward and other telescope resources will always include a mixture of modernized TCSes compatible with RTS2, and other systems including commercially available control software. It is therefore desirable for the network manager to be able to command, query, and receive responses from non-RTS2 observatories. Such an interface may also be useful in the event that, for example, one wishes to request an observation at a larger telescope that is not under control of the network.

The basic idea is that the manager distributes observations among observatories in a sensible way, listens to the status of individual telescopes, receives reports of completed observations, and returns data to an archive. It also needs to log and report the status of the network and of observatories so that a human operator can monitor the network. The functional requirements of the network manager are still being defined. These include technical requirements based on science cases (eg response time, schedule constraints) and how the manager will distribute observations among telescopes.

The next layer is the observatory control system. These are instances of RTS2 running at each telescope. Each observatory control system receives requests for observations, schedules them at its telescope, and communicates with the next layer down that directly controls the hardware. The observatory control system must also listen to weather and safety related monitors, to, for example, close the telescope for bad weather, or discontinue robotic operation if a safety interlock is triggered.  This layer is what we have been bringing into operation at the Kuiper 61'', and testing during extensive night-time observing in the past year.

The bottom layer includes the telescope and instrument control systems.  These are the telescope control system, TCS-NG, described below, and for the Mont4K camera at the 61'', the AzCam controller developed by the Steward Imaging Technology Lab. RTS2 talks to these systems through purpose-written drivers. The TCS-NG and AzCam are also the systems used by classical observers.
 
\section{Telescopes for ARTN}

\begin{figure} [ht]
   \begin{center}
   \includegraphics[height=10cm]{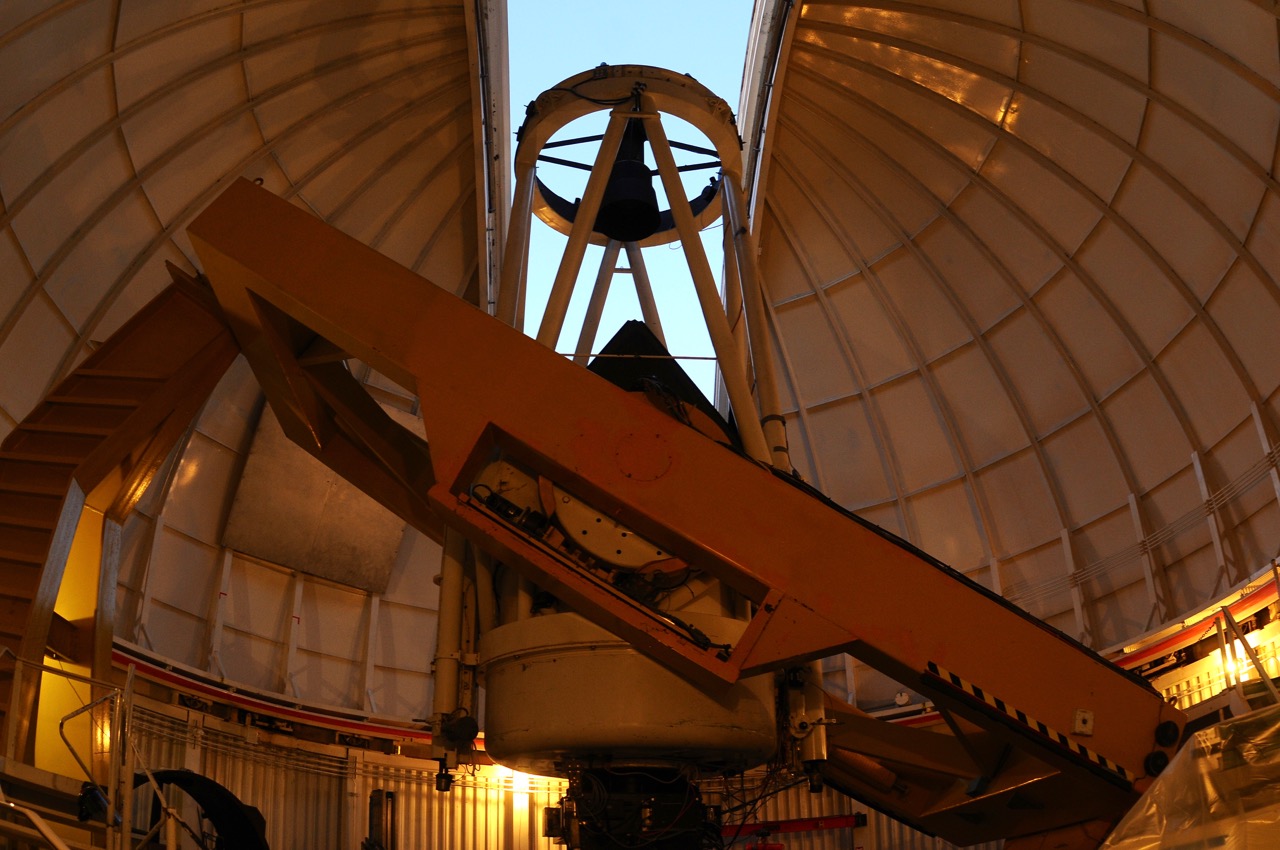}
  \end{center}
   \caption[] 
%>>>> use \label inside caption to get Fig. number with \ref{}
   { \label{fig:kuiper} 
The Kuiper 61'' telescope of Steward Observatory, with the Mont4K CCD camera mounted at the Cassegrain focus. The telescope is located on Mt. Bigelow, elevation 2510 meters, in the Santa Catalina mountains, one hour's drive from the University of Arizona.}
   \end{figure}

We have been using the Kuiper 61'' telescope, Figure \ref{fig:kuiper} as a rollout for the technology of queue scheduling, remote, and robotic operations.  The 61'' is located 
on Mt. Bigelow at 2510 meters elevation in the Santa Catalina mountains, a good site with relatively dark skies, yet just one hour's drive from the University of Arizona campus in Tucson. This makes frequent engineering much easier.  The Kuiper was originally built in the early 1960s to produce a photographic atlas of the Moon in advance of NASA lunar missions. Its control system has been upgraded over the years; the most recent upgrade is described briefly below. It has a modern 4Kx4K CCD camera with a 9.7\arcmin\ field of view. 

The Raymond E. White 21'' telescope in the original Steward Observatory dome on the U of A campus has also been used as a testbed for TCS development, although its use is primarily educational and will not be a regular part of ARTN.

The Vatican Advanced Technology Telescope (VATT) 1.8-m on Mt. Graham is also receiving upgrades to run the modernized TCS-NG described below, and will subsequently be tested with RTS2, either with a spectrograph or imager. A sophisticated ladder-logic interlock system has been designed to ensure the safety of people and property while operating the telescope without full-time on-site supervision.

Several of the small to medium Steward telescopes mostly used by the Catalina Sky Survey (Schmidt, Mt. Lemmon 40'' and 60''), can now run the new TCS-NG. These telescopes are frequently operated semi-remotely by CSS, e.g. one operator can run two telescopes.

The Bok 90'' telescope at Kitt Peak, which has a very wide field imaging camera, 90Prime, is running on Steward's older PC-TCS, but will eventually be converted to the new TCS-NG.

The SuperLOTIS 0.6-m telescope on Kitt Peak is already fully robotic, with an automated enclosure and weather monitoring, and carries out a nightly program of mostly transient observing. It runs on the older PC-TCS, but accepts a nightly observation specification in a simple ASCII form, and its data could potentially be integrated into the network.

Several additional small to medium telescopes operated by Steward or its partners are candidates for TCS upgrades and eventual integration into an ARTN.

%\section{Technical Challenges}

\section{System Upgrades}

\subsection{Telescope and Camera Hardware and Communication}

The Kuiper 61'' telescope underwent several hardware and software upgrades both as a matter of necessity and to update systems to allow for external control by the RTS2 control software.
The 61'' telescope right ascension and declination drives were upgraded to Copley Accelnet drive motor controllers. New and more robust limit logic was developed for hardware and software horizon limits.  The telescope is controlled by a computer running the QNX real-time operating system. The control system is based on communication via network sockets.

\begin{figure} [ht]
   \begin{center}
   \includegraphics[height=15cm]{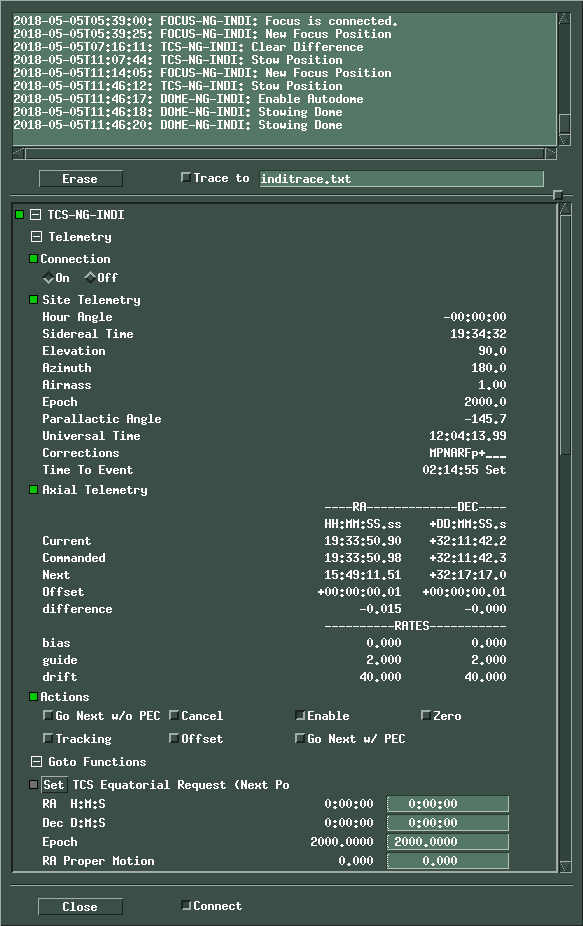}
  \end{center}
   \caption[] 
%>>>> use \label inside caption to get Fig. number with \ref{}
   { \label{fig:tcs-ng} 
A partial view of the TCS-NG INDI control panel window that allows the observer
to control the telescope and dome functions.
}
  \end{figure}

A new telescope control system, TCS-NG (TCS Next Generation), was developed by the Steward Observatory technical group to replace the previous Comsoft PC-TCS, which ran on obsolete PC hardware.  On the system side, TCS-NG uses sockets instead of the PC-TCS's serial communications, and is a newer, more maintainable and extensible codebase. On the user side, the PC-TCS used a text terminal and key-accessed menus; TCS-NG provides a Motif-based GUI for control of telescope motion, dome, and focus functions (Figure \ref{fig:tcs-ng}). Sections of the GUI are expandable to access various aspects of the telescope control, including special-use sections, such as non-sidereal bias rates. TCS-NG communicates via the Instrument Neutral Distributed Interface (INDI, {\tt http://www.indilib.org }), an open standard.

The TCS-NG runs together with an instance of the popular XEphem planetarium software,\\ {\tt http://www.clearskyinstitute.com/xephem/ }.  XEphem provides a view of the visible sky and shows the current telescope pointing and horizon limits. Object catalogs can be loaded into XEphem using common formats. These allow the user to select an object in XEphem from either its own catalogs or user catalogs, and send its coordinates to the TCS-NG process. A slew is then initiated by pressing a button in the TCS-NG window.  

The Mont4K camera CCD is controlled by the AzCam software written by the Steward Imaging and Technology Laboratory (ITL). This software runs on a Windows PC in the control room and can be operated classically, or operated by the RTS2 control system, using an RTS2 driver for AzCam written by P. Kub\'anek.  This will be extensible to Steward and other facilities using ITL detectors controlled by AzCam.

TCS-NG currently runs on the 61'', on the Raymond E. White 21'' telescope in the original Steward Observatory dome on campus, a useful local testbed, and on several telescopes primarily used by the Catalina Sky Survey. It is being implemented at the VATT and will eventually be rolled out to most of  Steward's small to medium telescopes, including the Bok 90'' at Kitt Peak.

\subsection{Control and User Software}

The RTS2 control software that communicates with TCS, camera, and other facility installations (e.g. weather) and commands the observations is in use at a number of telescopes.  It is written and maintained by P. Kub\'anek, is open source for Linux, and has been described extensively elsewhere\cite{Kubanek08,Kubanek16} and at {\tt https://rts2.org}. RTS2 is a complex and flexible engine with many control parameters.  In the engineering phase we are operating it hands-on, but everyday science users of the telescope system will not normally interact directly with it.  We have written Python scripts that can,  for example, translate a simple target list of observation descriptions (eg name, RA, Dec, and filter, exposure time pairs) into the RTS2 commands to generate observation plans and add them to a queue.  We are developing a web interface to take observation specifications. In the future this will become a user portal that handles both observation requests and returning data to the users.

We have begun tracking ARTN-related software and engineering through a github repository with an issue tracker and wiki. This has proven useful for opening, tracking and resolving issues, for archiving support scripts and recipes, and for collaborating on user documentation and how-to guides as the procedures for operating the queue are refined.

We have developed a simple Python pipeline for basic processing of the Mont4K CCD imaging.  The pipeline is based on {\tt ccdproc}, an Astropy-affiliated package\cite{astropy13,ccdproc15,astropy18}, and is intended to batch-process a night's worth of data, including science and calibration frames.
This pipeline generalizes ccdproc for basic reduction steps (overscan, trim, bias, flatfield) on multi-extension FITS files, since the Mont4K CCD has two readout amplifiers. It is not specific to Mont4K, but generic to FITS images with multiple extensions each representing observed flux, and thus can be used for a variety of CCD imagers and mosaics. The pipeline is released for public use at {\tt https://github.com/bjweiner/ARTN/tree/master/mont4k\_pipeline}. 

A local instance of the Astrometry.net software\cite{Lang10} ({\tt http://astrometry.net/}) runs on the 61'' observing computer and can be used to provide immediate astrometric calibration of images.  Gluing these reduction steps together to quickly produce calibrated images and return them to the science user is in progress.

\begin{figure} [ht]
   \begin{center}
   \includegraphics[width=15cm]{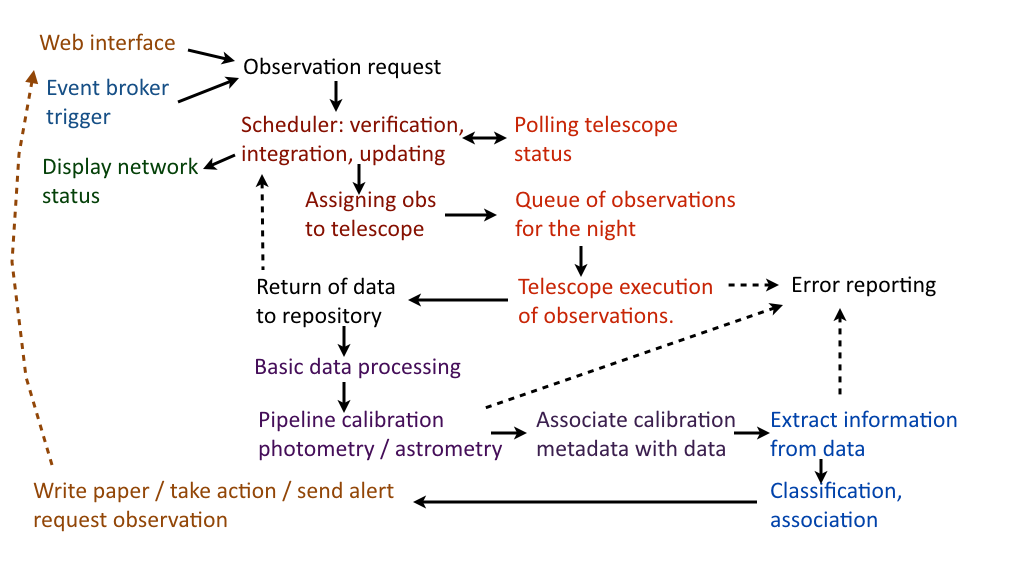}
  \end{center}
   \caption[] 
%>>>> use \label inside caption to get Fig. number with \ref{}
   { \label{fig:lifecycle} 
A schematic diagram of a suggested life-cycle for an observation, from trigger through request, scheduling, observation, data reduction, scientific processing, and return of data / information to the user.
}
  \end{figure}

The top layer of ARTN, the network manager, will require an interface to users to specify and trigger observations, and eventually a semi- or fully-automated interface to event brokers so that users can construct filters to select events and specify and trigger observations.  A potential life-cycle from observation to data is shown in Figure \ref{fig:lifecycle}.

  \begin{figure} [ht]
   \begin{center}
   \includegraphics[height=12cm]{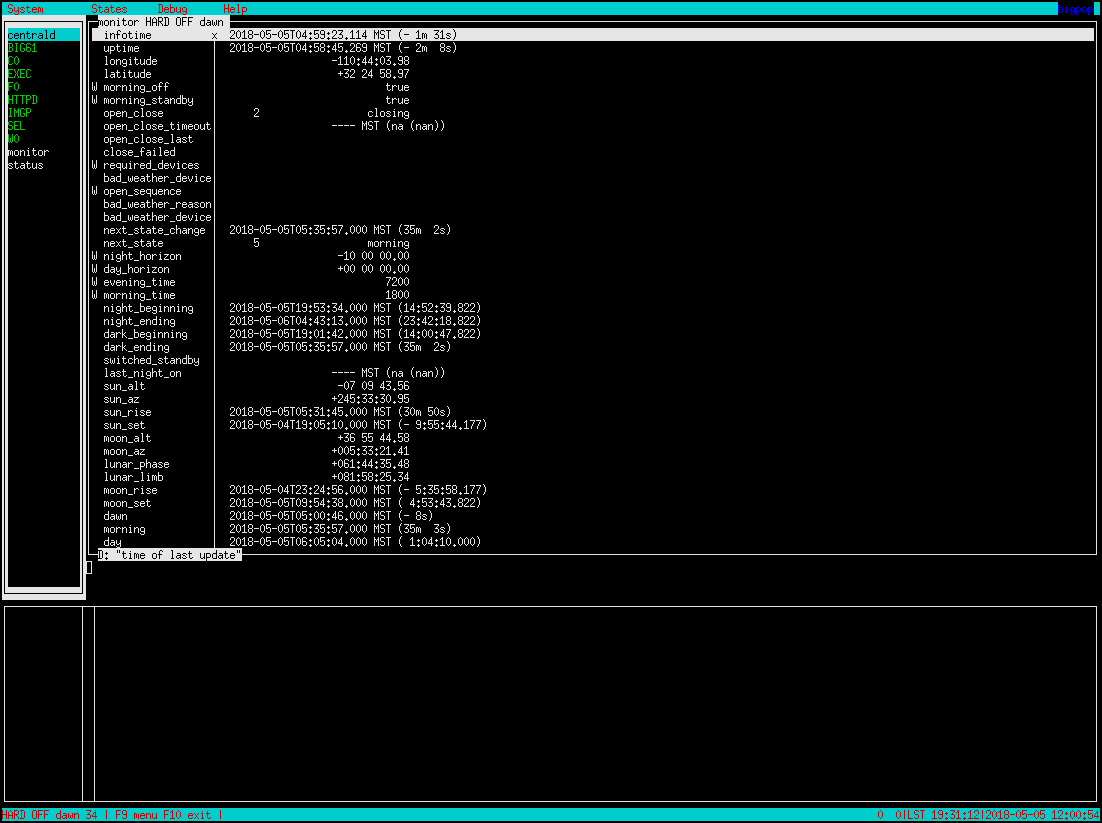}
  \end{center}
   \caption[] 
%>>>> use \label inside caption to get Fig. number with \ref{}
   { \label{fig:rts2mon} 
A view of the RTS2 monitor text interface window showing the main tab -- additional tabs selectable in the left frame describe telescope, camera, focus, and scheduler interfaces -- and a number of the parameters controlling the system behavior.
}
  \end{figure} 

\section{Operations and Safety}

Upgrading the TCS and developing the automated control at the 61'' has been done during engineering nights interspersed with the usual classically scheduled operations.  In addition, the Kuiper is sometimes used with instruments other than the Mont4K CCD camera (SPOL, and PI instruments), and some observing programs prefer classical to queue operations. Therefore classical and queue scheduling will co-exist at the 61'' for an extended time.  This limits the scope of tasks to implement remote/autonomous operations that can be carried out during any engineering run, since the telescope must remain operable in classical mode.  It also requires care to insure the safety of personnel and the facility.

The operation of an automated telescope and dome raises serious safety issues.  The telescope and dome must be capable of being locked-out when staff are or could be in the chamber. We must avoid the possibility of having either a remote observer or the automated control system wake up and move the telescope when people may be on the dome floor.
The control room of the Kuiper telescope is downstairs and is safe to enter and exit while systems are moving. However, it is currently necessary to access the dome floor to open and close the dome and mirror covers, and to fill the CCD dewar with liquid nitrogen.  Classical observers will not ordinarily command a telescope slew while they are themselves on the dome floor, and a wide field security camera is installed in the dome that allows an observer in the control room to watch the telescope while it slews during ordinary operations and notice unsual motions or obstructions.

An interlock has been designed to lock out remote control so that remote users or the automated system will be disabled and not allowed to move the telescope while personnel enter the dome. However, the interlock must also have a classical mode that allows classical or engineering users to enter the dome (with the usual care) without fully shutting down the telescope.
To date, testing of RTS2 control of the telescope has been done with experienced engineers or observers in the control room to watch it, and the interlock technology has not been activated. As queue scheduling/remote operation progresses out of the engineering phase, we plan to have an external safety review of the interlock system design before full implementation. 

Insuring safety of the facility and telescope also requires an interface to local weather monitoring and weather predictions, to shut down when conditions are outside the operating limits for wind or humidity, and to anticipate risk of precipitation and close conservatively.  A weather station and all-sky camera currently operate at Mt. Bigelow ({\tt https://www.lpl.arizona.edu/$\sim$css/bigelow/}). Developing conservative weather limits also awaits dome shutter automation; currently there must be an observer on-site. It is possible that in future, multiple telescopes on a single site will have a single minder.

\section{Current Status}

Currently, the Kuiper 61'' telescope and the Mont4K camera can operate under the control of RTS2 with full queue scheduling of nightime science observations, and automatically taking twilight sky flats. 
The queue scheduled observations currently being carried out are largely transient imaging, mostly of supernovae, and observation definitions (target, filters, exposure times) are generated by simple Python scripts that issue RTS2 commands.  A web interface allowing users to define observations is in development. Proposals from the Arizona user community for shared risk queue observing were invited for the 2018B semester. 
 
During this phase of testing observations, the telescope, camera, and queue are ``minded'' by an experienced observer, to set the system up each night, catch problems, and guarantee the safety of telescope and personnel.
At the moment the telescope must still be focused each night by an observer taking multiple exposures and measuring the stellar images; a routine to automatically take a focus sequence and determine the best focus is in active development.  The telescope dome and mirror covers are still opened and closed by the observer using a traditional control panel. Engineering upgrades are under way in summer 2018 to allow these systems to be controlled remotely.

The tracking and pointing of the 61'' are very good, with typical pointing better than 10\arcsec. The offset autoguider is not yet under RTS2 control, but for imaging observations, exposures are typically short enough that guiding is not required (several minutes).  For future robotized spectroscopy, guiding will be necessary.

A remote observing room at Steward Observatory's campus offices can be used to control the 61'' telescope. This facility has been tested, but since the dome shutter is not yet remotely controlled, it is not yet possible to operate the telescope fully remotely.  

The RTS2 queue scheduling of the telescope is presently being used in a simple order of object priority -- typically West to East order. In future, a more complex system of prioritizing observations and long-term scheduling will be added.

\section{Conclusions}

After some time of incremental development, the concept of the Arizona Robotic Telescope Network is beginning a scientifically usable data-taking phase with semi-automated operation of the Kuiper 61'' telescope and Mont4K camera under control of the RTS2 queue scheduler.  The system can operate the telescope and camera for a night of observations. Human supervision is currently still needed during this engineering phase, e.g. for opening and closing the dome, focusing, monitoring for image quality and system faults, LN2 dewar fills, and safety and weather checks. Many of these tasks are in progress and some may be doable by crew shared among multiple telescopes on these sites.  Our goals are to have the 61'' automated by end of 2018, the VATT by summer 2019, and the Bok 90'' by 2020.

However, the science goal of ARTN is to provide flexible and responsive observations, not to eliminate human supervision and decision making  from the network. Replacing the last small percentage of tasks performed by people, such as an experienced observer's decision making, is often the hardest. The goal of creating an autonomous telescope network should be to provide improved science capabilities, not to save money by removing observers, because that is not a realistic assessment of the labor involved.

Our experience with updating legacy telescopes to new hardware and software capable of remote or autonomous operation points to some of the difficulties or pinch-points for such a program. A major issue is the need to preserve the capabilities of telescopes that are in active use for classical observing. When a telescope is in demand and engineering time is limited to a few nights a month, progress can be slow, and the need to switch back and forth between old and new systems adds complexity and takes engineering staff time. It also limits the implementation of hardware such as safety interlocks. 

Stable funding for facility upgrade projects can also be an issue, as without dedicated funding, engineering staff are responsible for both upgrades and maintenance on multiple facilities. Although new facilities are attractive to funding sources, existing facilities on well-established sites are valuable to the astronomy community, especially given that new sites can be difficult to develop.
Finally, of course, the fact that a telescope is in demand for observing even while upgrades are in progress is not really a bad thing, as it shows that the facility continues to be scientifically relevant.

%The last n percent of automation is the hardest - may still want a person monitoring the network, overseeing multiple telescopes, or at least have the ability to have someone on call.

%%%%%%%%%%

% \appendix

\acknowledgments % equivalent to \section*{ACKNOWLEDGMENTS}       
 
We are grateful to the Papal Foundation, the Hearst Foundations, and the Dan Murphy Foundation for their support of VATT's upgrades and of the RTS2 code development; Steward Observatory and its Mountain Operations and Technical Support group for continued support of the ARTN project; and the University of Arizona Observatories for its support with substantial allocations of engineering and observing time.

% References
\bibliography{report} % bibliography data in report.bib
\bibliographystyle{spiebib} % makes bibtex use spiebib.bst

\end{document}